
\magnification=\magstep1
\baselineskip=14.25 pt plus 0.5pt minus 1pt
\vsize=23.5truecm
\hsize=16.0truecm
\voffset=-0.2truein
\hfuzz=1.5pt
\footline={\ifnum\pageno=1 \hfil\else\centerline{\folio}\fi}
\def\hi{\hangindent=15pt}
\def\vs{\vskip 6truept}

\def\veject{\vfill\eject}
\def\no{\noindent}
\def\h{\textstyle {1 \over 2}}
\def\sfrac#1#2{{\scriptstyle{#1\over#2}}}
\def\ft#1{${}^{#1}$~}
\def\p{\partial}
\def\setify#1{{$\{#1\}$}}
\def\E#1{Eq.~(#1)}
\def\Es#1{Eqs.~(#1)}
\def\ck{{\cal K}}
\def\ee{\hbox{\bf{e}}}

\font\smchapt=cmbx12
\font\smaller=cmr9

\font\smbold=cmbx8
\font\reglar=cmr10
\font\outline=msbm10
\font\hscript=eufb10 scaled \magstep 1
\textfont11=\outline
\textfont12=\hscript

\def\reel{{\outline R}}

\def\complex{{\outline C}}
\def\qa{{\fam=12 A}}
\def\ad#1{{(\hbox{ad}\,{\bf{ #1}})}}
\def\ads#1{{(\hbox{ad$\,${\smbold #1}})}}

\centerline{{\smchapt Estabrook-Wahlquist
Prolongations and Infinite-Dimensional Algebras}}\smallskip
\centerline{{\reglar J. D. Finley, III}}
\centerline{ Dept. of Physics and Astronomy, University of
New Mexico}
\rightline{Albuquerque, N.M., 87131 U.S.A.\qquad
finley@tagore.phys.unm.edu}
\bigskip
\reglar
\par\no{\bf I.~~Estabrook-Wahlquist Prolongations and Zero-Curvature
Requirements}\par
Motivated by a desire to find
new solutions of physically-interesting partial differential equations,
we think of a $k$-th order pde as a variety, $Y$,
of a finite jet bundle, $J^{(k)}(M,N)$, with $M$ the
independent- and $N$ the
dependent-variables for the pde.
 From this geometric approach, we can look for point symmetries,
contact symmetries,
generalized symmetries, or even non-local symmetries, where the system
is prolonged
further, to a fiber space over $J^\infty$, with fibers
$W$, where vertical flows map solution spaces of one pde
into another, satisfied by the additional dependent variables,
$w^A$, that coordinatize the fibers.  The compatibility
conditions for such flows to exist are {\it ``zero-curvature conditions.''}
Solutions of these
conditions may be found using the tangent structure
or the co-tangent structure, over $J^\infty\times W$.  We describe
both, but follow the approach via differential forms,
following Cartan,\ft{1}, Estabrook and Wahlquist,\ft{2} and Pirani,\ft{3}
believing that it provides better guides for the intuition, for
complicated (systems of) pde's.
\vs
For a vector-field presentation, we choose a commuting\ft{4} basis, $\{e_a\}$,
for tangent vectors over $M$, and lift them to the total derivative operators,
$D_a$,
over $J^\infty$. Provided the system of pde's is
involutive, they will still commute when restricted to the variety $Y^\infty$,
described by
the pde's, this restriction being denoted by $\overline D_a$.
The further prolongation into the fibers $W$ requires the addition of some
vector fields vertical with respect
to the fibers, i.e., $\b X_a = \sum X_a^A(\p/\p w^A)$,
with the $X_a^A$ functions of both the jet variables and
the \setify{w^A}.
Requiring that the $\overline D_a+\b X_a$ commute, when
restricted to $Y^{(\ell)}\times W$, for some $\ell$,
ensures that the $w^A$
can act as pseudopotentials for that pde:\ft{5,6}
$$0\ = \
\left[D_a+\b X_a\,,\,D_b+\b X_b\,\right]_{\big|_{Y^\infty
\times W}}\!\!\!\! =
\left\{\overline D_a(X_b^C) - \overline
D_b(X_a^C)\right\} {\p\over \p w^C}\ +\ \left[\b X_a\,,\,\b X_b\,
\right]\>.\eqno(1.1)$$
\no The general solution for the $\b X_a$
describes all
possible fiber spaces, or {\it coverings},\ft{1} for this pde,
As the construction gives the $\b X_a$ the ``form'' of a connection, it is
reasonable to refer to these equations as ``zero-curvature'' requirements; it
is, however, a generalization of the usual
approach,\ft{7,8} since the
$\b X_a$'s are still only elements of an {\bf abstract} Lie algebra of
vector fields, with neither coordinates, nor even their number yet determined.
\vs
 As an identity in the jet coordinates, \Es{1.1} determine
several independent equations.
Their solution describes the
$\b X_a$ as linear combinations of vector
fields $\b W_\alpha$ with coefficients depending on
coordinates for $Y^{(\ell)}$, with the
$w^A$-dependence encoded within a set of
commutation relations among the \setify{\b W_\alpha}, as
vector fields within the algebra of vector fields over $W$.  The
smallest subalgebra that faithfully reproduces the linear independence, and
the values, of those commutators
is the general solution to the covering problem; we believe it is a universal
object for the given pde and others
related to it, so that it may be used to characterize related classes of
pde's.\ft{6,9,10}
\vs   The isolation and identification
of such algebras is an important part of the
process of determining and understanding all the solutions
of nonlinear pde's.  Vector-field realizations will generate
B\"acklund transformations, inverse scattering
problems, etc.\ft{10,11}  Faithful
realizations will usually involve infinitely many pseudo-potentials, making
their identification somewhat difficult, and the first researchers
did not consider the entire
infinite-dimensional algebras.  However, beginning
with the work by van Eck,\ft{12} and Estabrook,\ft{13} on
identification of the
universal algebra for the KdV equation, the search for the
infinite-dimensional algebras involved has been extended considerably
by Hoenselaers and co-workers\ft{10,14,15}, by Omote\ft{16},
and by the group at
Twente, who seem to have made this a studied art-form.\ft{17}
\v
The dual approach,
via differential forms, created by Estabrook and Wahlquist
and built on the ideas of Cartan,
begins with the `contact module,'
 $\Omega^k(M,N)\subseteq [J^k(M,N)]_*$, generated by the
following set of 1-forms:
$$\smaller \Omega^k (M,N):\! \left\{ \matrix{ \theta^\mu = \,dz^\mu - z^\mu_a d
x^a ,\hfill \cr
\qquad\qquad\quad\ldots \hfill\cr
\theta^\mu_{a_1a_2 \ldots a_{k\!-\!1}}\!\! = \,dz^\mu_{a_1a_2 \ldots
a_{k\!-\!1}} \!\!-z^\mu_{a_1
a_2 \ldots a_{k\!-\!1} a_k} dx^{a_k}\hfill\cr } \right\}
\equiv \left\{\theta^\mu_\sigma\mid |\sigma| = 0,
\ldots, k-1\right\} \eqno(1.2)$$
\no where the summation convention is being used, and a choice for a local
coordinate chart is
$\left\{x^a,z^\mu, z^\mu_a, z^\mu_{a_1a_2},\ldots, z^\mu_{a_1 \ldots a_k}
\right\}$, with $z^\mu_\sigma$ standing for any of these
(jet) coordinates except the independent variables themselves, $x^a$.
The contact module `remembers' the relation
the coordinates of the jet bundle would have when they are pulled
back by the lift of
a function over $M$: $u:U\subseteq M\rightarrow N\>\Longrightarrow\>
(j^ku)^*(\Omega^k) = 0$.  The ideal, ${{\cal I}}$, is the
differential closure of the
pullback of the contact module to $Y$, and constitutes
the Cartan description of the original pde.
For 2 independent variables, \setify{x,y}, the EW procedure
first chooses a {\bf proper}, closed subideal,
${\cal K}\subset{\cal I}$, generated by a set of
{\bf 2-forms}, \setify{\alpha^r},
that still is effective at describing the given pde.\ft{18,19}
The new variables \setify{w^A} are adjoined by appending to
$\cal K$ contact forms, \setify{\omega^A}, for each of these pseudopotentials,
and maintaining the ideal closed:
$$\eqalign{\omega^A  =  & - dw^A + F^A dx + G^A dy, \quad\cr
dF^A \wedge dx + & dG^A\wedge dy = f^A\!\!{}_r\, \alpha^r + \eta^A{}_B
\wedge \omega^B\quad,\cr}\qquad  A = 1, \ldots, N\quad.\eqno(1.3)$$
\vs To show ``equivalence'' with
the zero-curvature equations, we first consider all of $J^{(k)}$, i.e.,
without a pde, and then ``restrict'' to $Y$.
For simplicity considering quasi-linear pde's,
we first select $\cal L$, generated by the wedge-products of
 all 1-forms in $\Omega^k$
 with the  $dx^a$. For $0 \le |\sigma| \le k-1$, $\cal L$, contains exactly
one copy of each of $dz^\mu_\sigma$.  Labelling its \veject
\no coefficient, in \Es{1.1},
by ${(f^A)}^a_{z^\mu_\sigma}$, we have
$$F^A_{,z^\mu_\sigma} = {(f^A)}^1_{z^\mu_\sigma}\quad,\quad
G^A_{,z^\mu_\sigma} = {(f^A)}^2_{z^\mu_\sigma}\quad, \eqno(1.4)$$
\no for each jet coordinate $z^\mu_\sigma$, with no repetitions.
Writing $(\eta^A{}_B)_{z^\mu_\sigma}$ for the
components of the 1-forms $\eta^A{}_B$,
\Es{1.4} also gives us
$$\eqalign{ 0 = (\eta^A{}_B)_{w^C}\, dw^C\wedge dw^B\;,&\quad
 0 = (\eta^A{}_B)_{z^\mu_\sigma}dz^\mu_\sigma\wedge dw^B\;,\cr
F^A_{,w^B}dw^B\wedge dx = -(\eta^A{}_B)_x dx\wedge dw^B\,,&\quad
G^A_{,w^B}dw^B\wedge dy = -(\eta^A{}_B)_y dy\wedge dw^B\,,\cr
\Longrightarrow\qquad\eta^A{}_B = F^A_{,w^B}dx + &
G^A_{,w^B}dy\quad.}\eqno(1.5)$$
\no The only remaining part of \Es{1.4} are the coefficients of the
basis 2-form $dx\wedge dy$:
$$-F^A_{,y} + G^A_{,x} = -F^B\,G^A_{,w^B} + G^B\,F^A_{,w^B} +
\sum_{|\sigma| = 0}^{k-1}\left\{-z^\mu_{\sigma y}F^A_{,z^\mu_\sigma} +
z^\mu_{\sigma x}G^A_{,z^\mu_\sigma}\right\}\quad.\eqno(1.6)$$
Introducing vertical vector fields, ${\bf F}\equiv (F^A)\p/\p w^A$ and
${\bf G}\equiv (G^A)\p/\p w^A$, so that the first two terms on the
right hand side are the components of a commutator, this becomes
$$[D_x + {\bf F}, D_y + {\bf G}] =
[\p_x + {\bf F},\p_y + {\bf G}] + \sum_{|\sigma| = 0}^{k-1}\left\{
-z^\mu_{\sigma y}{\bf F}_{,z^\mu_\sigma} +
z^\mu_{\sigma x}{\bf G}_{,z^\mu_\sigma}\right\} = 0\quad.\eqno(1.6^\prime)$$
\vs This has the same form as \Es{1.1}, except that we must still effect
the restriction to some variety $Y$. The resulting EW ideal, $\cal K$,
will be defined over $Y^{(k-1)}\equiv Y\cap J^{(k-1)}$, so that
it reduces the problem in an
important way; $\bf F$ and $\bf G$ will depend on
several fewer variables---only
the coordinates for $Y^{(k-1)}$, which we select by choosing
 ``co-coordinates'' for $Y$, i.e., those jet
coordinates the pde's will be used to eliminate, as a method of
(locally) defining $Y\subset J^{(k)}$.\ft{20}  If the restriction of
 $\cal L$ to $Y$
removes all the $k$-th level coordinates, then it can be taken as $\cal K$.
Otherwise, the remaining highest derivatives
must still be removed from the system, which is always possible,
although the methods depend on the particular
pde. For a simple, quasi-linear evolution equation, we choose
$z_y = H\,z_{(k)} + K$ as our co-coordinate, where $H$ and $K$ are functions
over $Y^{(k-1)}$. Restriction to $Y$ then causes $z_{(k)}$ to appear twice:
in $(dz - z_y\,dy)\wedge dx$ and also $(dz_{(k-1)} - z_{(k)}\,dx)\wedge dy$.
The following replacement
process, followed by dropping the second 2-form above,
reduces $\cal L$ to $\{Y^{(k-1)}\}^*$,
as desired:
$$\eqalign{(dz  & - z_y\,dy)\wedge dx  \rightarrow \left\{dz -
(H\,z_{(k)} + K)\,dy\right\}\wedge dx \cr \equiv &
dz\wedge dx + H\,dz_{(k-1)}\wedge dy - K\,dy\wedge dx\ \hbox{\rm mod}
\ (dz_{(k-1)} -
z_{(k)}\,dx)\wedge dy\,.\cr}\eqno(1.7)$$
\vs
For an evolution equation, the above process is unique,
and the EW process gives
exactly the same results as that using vector fields; however,
in general the situation is different.  As a second example, we
consider a pde
that defines $Y$ via the equation $z_{yy} = H\,z_{(k)} +
J\,z_{(k-1),y} + K$, where the integers in parentheses indicate the
number of $x$-derivatives, and
$H$, $J$, and $K$ are defined over $Y^{(k-1)}$.
The earlier replacement process now {\bf has more than one path
allowed}:$$\openup+1\jot\eqalign{(dz_y -  z_{yy}&\,dy)\wedge dx\quad
\longrightarrow
\  dz_y\wedge dx - K\,dy\wedge dx + H\,dz_{(k-1)}\wedge dy\qquad\cr &\quad +
J\,\cases{
dz^\mu_{(k-2),y}\wedge dy\,,& Option 1\cr
- dz^\mu_{(k-1)}\wedge dx\,,& Option 2 (equiv. to 1)\cr
\h (z^\mu_{(k_a-1)}\wedge dy - z^\mu_{(k_b -1)}\wedge dx)
\,,& Option 3, symmetric, \cr
& \quad but inequivalent.\cr}}\eqno(1.8)$$
\no The two inequivalent paths each generate
acceptable EW ideals, but lead to distinct prolongation structures.
A third example pde with inequivalent prolongation structures
is the sine-Gordon equation, as described below.
\vs
We describe the generators in $\cal K$ as a set of contact
2-forms for only those coordinates needed
for $Y^{(k-1)}$, {\bf and}
 a (set of) ``dynamical 2-forms'' for each pde in the
system.  The ideal $\cal K$ provides us
a geometrically-motivated structure for knowing on which jet-variables we
need no dependence; as well, the Lagrange multipliers,
$f^A{}_r$, expressed
in terms of derivatives of the $F^A$ and $G^B$, as in \Es{1.4}, tell us on
which
of the coordinates they {\bf must} depend.
(The curvature should vanish {\bf only} when it is restricted to
$Y$, so that the $w^A$ are truly pseudopotentials for the pde;
within the functional form of the curvature, the $f^A{}_r$
multiply the 2-form expression of the pde. The remaining information is
then the commutator equation, \E{1.6${}^\prime$}.
\vs\vs
\no{{\bf II.~~$\bf u_{xy} = f(x,y;u):$ The sine-Gordon and
Robinson-Trautman equations}}\par
Using the generalized form of the equation, with any choice of $f$,
we begin with the ideal, $\cal L$,
as just described.  Using the pde
to replace $u_{xy}$ within $\cal L$ leaves us with the largest possible EW
ideal, with 4 generators.  This ideal generates exactly
the vector-field commutator equations,\ft{6} and is too large
to allow us to solve the resulting equations.  However, Pirani,\ft{3} has shown
the existence of two distinct, useful sub-ideals:
$${\cal K}_1:\ \left\{\eqalign{(du - p dx)&\wedge dy \cr
(dp-f\,dx)&\wedge dx \cr}\right\}\ ,\qquad
{\cal K}_2:\ \left\{\eqalign{(du - p\,dx)&\wedge dy \cr
(du - q\,dy)&\wedge dx \cr  dp\wedge dx - dq\wedge& dy + 2\,f\,dx\wedge dy\cr}
\right\}\ .\eqno(2.1)$$
\v
For the case $f=f(u)$, only, such as the sine-Gordon equation, we have
$$\eqalign{ [{{\bf F}} \,,\,{\bf G}] = -p\,{\bf G}_u & +q\,{\b F}_u + f(u)
({\b F}_p-{\bf G}_q)\;,\cr
\hbox{with }\ {\b F} = {\b F} (u,p;w^A)& \;,\quad
{\bf G} = {\bf G}(u,q;w^A)\;,\cr
\hbox{{\bf and}}\quad\ck_1:\quad
{\b F}_u = 0 = {\bf G}_q & \;, \qquad\ck_2: \quad {\b F}_p + {\bf G}_q = 0
\ .\cr} \eqno(2.2) $$
These equations are of quite a different character than those
for evolution equations.  For an evolution equation, the constraints
resolve the dependence over $Y$ by the solution of algebraic equations;
however, for higher order equations,
vector-field-valued pde's must be solved,
substantially increasing the
difficulty of the problem.\ft{21}
Nonetheless, with relatively minor assumptions, the reductions
generated by these smaller algebras allow us to resolve
these equations as the ``flow'' of one vector field along the direction
described by another, and to express the solutions
explicitly in terms of the adjoint operation of one field upon
another.\ft{22} We describe the general solutions of each,
labelling the resultant algebras by
$\qa_1$ and $\qa_2$, and the prolongations $\{\bf F, G\}$ within them as
$\{{\cal F}_i, {\cal G}_i\mid i=1,2\}$.  We also show that $\qa_2$ is gauge
equivalent to a subalgebra of $\qa_1$, and that a subalgebra of that is
homomorphic
to $A_1^{(1)}$.  Important earlier work on
{\it infinite} versions of these
algebras was done by Hoenselaers,\ft{10,15} by Leznov
and Saveliev,\ft{8} by Dodd and Gibbon\ft{23}
for ${\cal A}_1$, and by Shadwick,\ft{24} for ${\cal A}_2$.
\vs
With details in Ref. 22, the solution for
$\qa_1$ may be found by first defining ${\bf Z}\equiv {\cal G}_1+({\cal
G}_1)_{uu}$, which requires
$[{{\cal F}}\,,\,{\bf Z}] = - p\>{\bf Z}_u$.  Expanding
${{\cal F}}$ about the origin and writing ${\bf F}_n$ as the
coefficient of $p^n/n!$, the flow equations tell us that
$${\bf Z} =\ee^{-u\>\ad{{\b F}_1}}{\bf Q}_0\quad,\qquad[{\b F}_n\,,\,
\ad{{\bf F}_1}^m{\bf Q}_0]
= 0,\quad,\quad n\ne 1\,,\, m = 0,1,2,\ldots\quad,\eqno(2.3)$$
\no where ${\bf Q}_0\in W_*$ is a ``constant'' of the integration.
Continuing in this mode, the general solution is
$$\eqalignno{  {\cal F}_1 & - {\b F}_0 + p{\b F}_1 =
\int_0^p ds\, (p-s)\, e^{s\>\ad{{\bf G}_1}}  {\bf K}_0
= \sum_{n=0}^\infty {p^{n+2} \over
(n+2)!}\ad{{\bf G}_1}^n {\bf K}_0\,,\cr
{\cal G}_1 - & {\bf G}_0 \cos u - {\bf G}_1 \sin u = \int_0^u
dw \sin (u-w) e^{-w\>\ad{{{\b F}_1}}}{\bf Q}_0
=  \sum_{n=0}^\infty\,(-\cos u)^{(-n-2)}\ad{{\bf F}_1}^n{\bf Q}_0\cr}$$
\no and requirements on {\bf (only)} some of the commutators:
$$\eqalignno{[\b f_0,\b g_0]=0\,,\;[\b f_0,\b g_1] =
-\b f_1\,,\;[\b f_0,\b q_n] = 0\,, & \;
[\b f_1,\b g_0]=\b g_1\,,\;[\b f_1,\b g_1] = -\b g_0 +\b q_0-\b k_0\,,\cr
[\b f_1,\b q_n]=-\b q_{n+1}\,,\;[\b k_m,\b g_0] = 0\,, & \;[\b k_m,\b g_1] =
-\b k_{m+1}
\,,\;[\b k_m,\b q_n] = 0\cr
{\bf K}_n \equiv \ad{G_1}^n\;{\bf K}_0\quad,&\qquad{\bf Q}_m \equiv
(-1)^m\ad{F_1}^m\;{\bf Q}_0 \;.\cr}$$
\vs
Alternatively, the equations that define
$\qa_2$ first tell us that
$$\qa_2 : \left\{\matrix{{{\cal F}_2} = \h p{\bf R} +
{\bf B}\quad,& \quad{{\cal G}_2} = - \h q {\bf R} + {\bf C},& \quad {\bf R}_{u}
= 0\quad,
 &\cr\cr \left[ \h {\bf R}\,,\, {\bf B}\right]  = {\bf B}_{u}\quad,&
\quad \left[ \h {\bf R}\,,\, {\bf C}\right] = - {\bf C}_{u}\quad,& \quad[{\bf
B}\,,\,
{\bf C}] = {\bf R}\,f(u). &\cr }\right.\eqno(2.4) $$
\no Integration of the differential equations gives
 two new, vertical vector fields such that
$${\bf B} = e^{+{{\scriptscriptstyle{1 \over 2 }}}
u(\ad {\bf R})} {\bf E}\;, \quad {\bf C} =
e^{-{{\scriptscriptstyle{1 \over 2 }}} u(\ad {\bf R})}
{\bf J}\quad.\eqno(2.5) $$
\no  Defining iterated commutators,
${\bf E}_m \equiv
(+\hbox{ad}{\bf R})^m{\bf E}\,,\;{\bf J}_n \equiv (-\hbox{ad}{\bf R})^n{\bf
J}$,
and setting $c_n$ as the coefficients multiplying $u^n/n!$
in the power series for $f(u)$,
the content of the commutator equation becomes simply
$$\bigl[{\bf E}_{k-m},\,{\bf J}_m\bigr] = c_k\,{\bf R}\;,\quad\forall\>m \ni
0\le m\le k \quad.\eqno(2.6) $$
\vs Even when divided by the countably infinite set of relations in \Es{2.6},
the free algebra generated by \setify{\bf J, R, E},
is so far unidentified as an already-studied algebra.
In fact, those relations do not appear to be consistent with the
usual sorts of gradings, so that
some distinct approach to infinite-dimensional algebras may be required.
We mention two quite different avenues that can be followed at this point.
If one no longer requires all the $\b E_m$ to be linearly independent,
then the homomorphic mapping to $A_1^{(1)}$ can be
demonstrated.  Alternatively, maintenance of linear independence
seems to lead in the direction of Toeplitz algebras of operators over
Banach spaces.
\vs
Viewing $\bf F$ and $\bf G$ as the components of a
 Lie-algebra-valued connection 1-form, $\Gamma$, over the covering spaces,
a gauge transformation generated by a vertical vector field, say ${\bf S}$,
would transform $\Gamma$ by
$\Gamma_t \equiv e^{t(\ad{\bf S})}\Gamma - d(t\,{\bf S})$.
Choosing $\b S = -\sfrac{1}{2}{\bf R}$, transforms $\qa_2$
into that part of $\qa_1$ generated by setting $\b K_0 = 0$.
$$\Xi:\,(\Phi_{-{\scriptscriptstyle{1\over 2}}u})_*(\qa_2)
\rightarrow(\qa_1)_{\big|_{{\bf K}=0}}\!\!\!
:\,\left\{\eqalign{{\bf R}\rightarrow {\bf F}_1
\,,&\;\b E_0\rightarrow{\bf F}_0\,,
\;\b J_0\rightarrow {\bf G}_0\,,\cr
\b J_1\rightarrow  & -{\bf G}_1\,,\;\dots\,,\;
\b J_j\rightarrow \;{\bf Q}_{j-2} - \b J_{j-2}\,.\cr}\right.\eqno(2.7)$$
Hoenselaers'
homomorphism\ft{10} of part of $\qa_1$ into $A_1^{(1)}$ can now be
extended to $\qa_2$:
$$ \eqalign{{\bf E}_m \equiv & \ad {\b R}^m{\bf E}\
\mathop{\longrightarrow}_\Xi
\ \ad {{\bf F}_1}^m\b F_0\ \mathop{\longrightarrow}_\Psi\
\cases{(-1)^{{m-2\over 2}}\>J_1^{(1)},& for m even,\cr
(-1)^{{m-1\over 2}}\>J_2^{(1)},& for m odd,\cr}\,,\cr
[\b E_m,\b E_n]\  & \ \mathop{\longrightarrow}_\Xi
\ [\ad {{\bf F}_1}^m\b F_0\,,\,\ad {{\bf F}_1}^n\b
F_0]\mathop{\longrightarrow}_{\Psi'}\ \cases{0\,,& $m + n$ even,\cr
(-1)^{{m+n-1\over 2}}\,J_3^{(2)}\,,& $m + n$ odd.\cr}\,.\cr}$$
\no
Triple
commutators of the ${\bf E}_m$'s among themselves will generate elements
of $A_1^{(1)}$ at the third level, etc.  While interesting, this homomorphism
loses much information contained within the larger algebra; for example,
$\Psi'(\b E_{m+2}) = -\Psi'(\b E_m)\,,\Psi'(\b J_{k+2}) = -\Psi'(\b J_k)$, and
it
eliminates any information carried by the generators
$\b Q_i$ and $\b K_j$.
\vs A quite distinct approach would maintain linearly independent at least
 those of the generators that appear in ${\cal F}_2$ and ${\cal G}_2$, which
leads us to consider that subalgebra
spanned, as a vector space, on the countable
list of generators \setify{{\bf R}, {\bf E}_m,
{\bf J}_n \mid m,n = 0,1,\ldots}, therefore requiring
the double, triple, etc. commutators
to be linear combinations of these, such as $[\b E_m,\b E_n] =
\left(A^i{}_m\right)_n\,J_i$.  This is a quite distinct approach from that
which led to $A_1^{(1)}$, where
only $\{\b R,\b E_0,\b E_1, \b J_0, \b J_1\}$, of this original set, were
linearly
independent, and it was their commutators that generated the higher levels of
the
Kac-Moody algebra.  Solving the constraints on the coefficients determining the
linear combinations leads to the creation\ft{22} of a
Banach algebra of Toeplitz operators made from countable sums of those
coefficients that define the commutators.  This mode of thinking causes us to
re-describe the countable set of repeated commutators, $\b E_m$, in terms of a
function $\b E(t)$, defined within a
Banach spaces of functions on the circle,\ft{25} along with a set of integral
equations, reminiscent of Weiner-Hopf equations,\ft{25} for functions defined
on
$S^1\times S^1$.
The resulting prolongation forms,
${\cal F}_2$ and ${\cal G}_2$, would then be expressed as integrals over that
circle:
$${\cal F}_2 = {\h} p{\bf R} -{1\over 2\pi i} \oint {dt\over t}\;\b E(t)
e^{{1\over 2} ut}\quad,\quad
{\cal G}_2 = -{\h}q{\bf R} -{1\over 2\pi i}\oint{dt\over t}\;\b J(t)e^{-{1\over
2} ut}\quad\eqno(2.8)$$
\vs
We may now return to the case where $f$ does depend on the independent
variables; an important example for this case is
the Robinson-Trautman equation for diverging, non-twisting,
Petrov type III solutions of the vacuum Einstein field equations:
$$\hbox{RT equation of type III:~~} u_{xy} =  \h(x+y)e^{-2u}\quad.\eqno(2.9)$$
\no Using the symmetric subideal, ${\cal K}_2$, in \Es{2.1},
the RT version of \Es{2.2} is
$$\eqalign{{\bf F} = p\,{\bf Z} + {\bf B}\quad,\quad{\bf G} & = - q \,{\bf Z} +
{\bf C}\quad,\quad
{\bf Z}_u = 0\quad,\cr
[{\bf Z}\,,\,{\bf C}] = -{\bf C}_u + {\bf Z}_y\quad,&\quad
[{\bf Z}\,,\,{\bf B}] = +{\bf B}_u + {\bf Z}_x\quad,\cr
[{\bf B}\,,\,{\bf C}] = {\bf B}_y-{\bf C}_x & +
(x+y)e^{-2u}{\bf Z}\quad.\cr}\eqno(2.10)$$
\vs   Since the sine-Gordon equation is well-studied,
we were interested in ``all'' of its subtleties.  On the other hand,
no interesting solutions of the RT equation are known; therefore we will
look, first, for as simple a solution as possible.
Obviously the prolongation structure must depend on $\{x,y\}$.
We have considered two complementary cases in Ref. 26, integrating there these
three-term pde's that generalize our earlier flow equations.  We also show
there that
(at least) subalgebras for each case are gauge equivalent.
Here, we only follow the simplest case, which assumes that
${\bf F}_y = 0 = {\bf G}_x$, thereby requiring that
$\bf Z$ be independent of both $x$ and $y$, and reducing our pde's
to simple flows, which integrate as before.  The additional assumption
that all commutators of $\bf Z$ with their (respective) constants of
integration
are parallel gives us a system quite similar to the generators
for {\bf sl}(2), but with $\{x,y\}$-dependence:
$$\eqalign{{\bf B}(x,u) = e^{+u\ads{Z}}\,{\bf R}(x)\qquad,&\qquad
{\bf C}(y,u) =
e^{-u\ads{Z}}\,{\bf S}(y)\quad,\cr
{\bf [ Z,\, S] = S\quad,\quad [Z,\,R] = - R}\quad,&
\quad \Longrightarrow\quad
[\b R(x),\,\b S(y)] = (x+y)\,{\bf Z}\quad.\cr}\eqno(2.11)$$
\vs As the independent variables occur linearly, the simplest,
non-trivial solution is for $\b R$ and $\bf S$ to be linear polynomials,
which causes the entire EW prolongation algebra to have a contragredient form:
 $$\eqalign{{\bf R}(x) \equiv -{\bf f}_1 - x{\bf f}_2\qquad,\qquad{\bf
S}(y)\equiv
+{\bf e}_2 + y{\bf e}_1\quad,\cr
\noalign{\vs}
[{\bf Z},{\bf e}_i]={\bf e}_i \quad,\quad[{\bf Z},{\bf f}_i]
=-{\bf f}_i\quad,\;i=1,2\quad,\cr
[{\bf e}_2,{\bf f}_1] = 0 = [{\bf e}_1,{\bf f}_2]\quad,
\quad [{\bf e}_1,{\bf f}_1] = {\bf Z} =
[{\bf e}_2,{\bf f}_2]\quad.\cr}\eqno(2.12)$$
\vs The three Lagrange multipliers are now
proportional to $\{{\bf R}(x), {\bf S}(y), {\bf Z}\}$.  We must therefore
determine
a realization of the algebra defined by these 5 generators,
which maintains these three linearly independent.
At some level, this is a simple task, since this is in fact just the
simplest contragredient algebra of infinite growth, usually referred to as
$K_2$,\ft{27} when one identifies our $\bf Z$ with its generator $h$.  The
generic contragredient algebra\ft{27} has the (standard) form
$$[{\bf e}_i,{\bf f}_j] = \delta_{ij}{\bf h}_i\quad,\quad [{\bf h}_i,
{\bf h}_j] = 0\quad,\quad [{\bf h}_i,{\bf e}_j] =
A_{ij}{\bf e}_j\quad,\quad [{\bf h}_i,{\bf f}_j] =
-A_{ij}{\bf f}_j\quad,\eqno(2.13)$$
where the $A_{ij}$ are elements of a
(generalized Cartan) matrix A.  Our algebra $K_2$
is just the algebra described
by the matrix $A = {~1~1\choose~1~1}$, divided by its center, $h_1-h_2$.
However, since no realizations of
$K_2$ have yet been discovered, this is not the end of the task.
\vs
The infinite growth of $K_2$ is of course the difficulty
To see how it affects the problem directly, we note that the
Jacobi identity requires
$[{\bf Z},\{\ad{e_1}^n{\bf e}_2\}] = (n+1)
\{\ad{e_1}^n{\bf e}_2\}$, and also
$[{\bf f}_2,\{\ad{e_1}{}^{m+1}{\bf e}_2\}]
 = -\h m(m+1)\{\ad{e_1}^m{\bf e}_2\}$.
The first equality shows us the growth of the dimension of the space of
$i$-th level commutators, {\bf unless} the objects $\{\ad{e_1}^n{\bf e}_2\}$
were to vanish from some value of $n$ onward. The second equality
tells us that this would cause a downward cascade,
leaving us with zero values for our Lagrange multipliers.
To give precise definitions, we define an (integer)-graded Lie algebra
as one that can be presented as
a direct sum of subspaces, which
the Lie bracket operation ``preserves''; i.e., for ${\cal G} =
\oplus{\cal G}_i$, we have
$[{\cal G}_i\,,\,{\cal G}_j] \subseteq {\cal G}_{i+j}$.
\no If $d_n$ is the dimension of $\sum_{j=-n}^n{\cal G}_j$, then the
(Gel'fand-Kirillov) growth,\ft{28}
$r$, of $\cal G$ is defined as
$\overline{\lim}_{n\rightarrow\infty}\left\{{\log d_n/\log n}\right\}$.
For $K_2$ one
finds that $d_n$ grows like $2^n$, so that the resulting growth is infinite.
 More recently, Kirillov\ft{29} has introduced the notion of algebras of
intermediate growth, where $\log d_n$ grows like $n^\delta$, for some
$0 < \delta < 1$, and has shown that $Vect(\reel^m)$ is
an algebra of intermediate growth, with $\delta = m/(m+1)$.  We conclude
from this that $K_2$, which has $\delta=1$, will not have a
realization within $Vect(\reel^m)$ for any finite value of $m$.
\vs
Nonetheless, the next step in
the process of finding new solutions is to determine explicit
realizations of this algebra, use the variables in
the carrier space as pseudopotentials, pick out a B\"acklund
transformation, take the one existing solution, and begin to generate
new ones, as has been done many times before with many other
pde's.  Since this is indeed the minimal
prolongation algebra, it seems reasonable to suppose that finding
new solutions is equivalent to evolving realizations of this algebra.
We hope to encourage
listeners to achieve a faithful realization of $K_2$.
\vs
\no{{\bf III.~~Even More Complicated Vector-Field PDE's}}\par
Quite an interesting system of vector-field-valued pde's has recently
arisen in investigations of a student, Denis Khetselius, who originally
came to UNM from Dubna.  His work on  twisting, Petrov type N vacuum
solutions lead him to the following (involutive) set of
{\bf coupled} equations:
$$\eqalign{{\cal L}_j{\cal E}_i - {\cal A}_i{\cal M}_j  & = [{\cal E}_i,{\cal
M}_j]\;,
\quad \forall i,j = +,0,-\quad,\cr
{\cal L}_+ \equiv a\p_b - e\p_f\,,\;
{\cal L}_- \equiv b\p_a - f\p_e\,, & \;
{\cal A}_+ \equiv u\p_w\,,\;
{\cal A}_- \equiv w\p_u\,,\;\cr
{\cal L}_0 \equiv [{\cal L}_+,{\cal L}_-]\,,&\quad
{\cal A}_0 \equiv [{\cal A}_+,{\cal A}_-]\,,} \eqno(3.1)$$
\no where the vector fields ${\cal M}_j$ and, separately, the
${\cal E}_i$ generate a realization of {\bf sl}(2,\complex) in
their (pseudopotential-type) variable spaces.  Notice, of course, that
the differential operators ${\cal L}_j$ and ${\cal A}_i$ also constitute
realizations of the generators for {\bf sl}(2,\complex).
(The quantity $s\equiv af+eb$ is a characteristic for
all the ${\cal L}_i$.  If one treats the 4 variables as complex, projective
coordinates, in \complex$^4$, for the group manifold, $s$ is the radius
variable.)
\vs
We can write the most general solution to this system of pde's, which is
quite ``messy.''  It can be done in a
number of different, but equivalent, ways, depending upon the order of the
integrations performed.  However, one could hope for a much less
coordinate-dependent approach to such a problem.
To emphasize the meaning of this quest, we
first study a slightly reduced version, obtained by assuming the
${\cal M}_j$ to be constant:
$${\cal L}_j{\cal E}_i = [{\cal M}_j,{\cal E}_i] = \{\hbox{ad}\,{\cal M}_j\}
{\cal E}_i\;.\eqno(3.2)$$
These equations may be
treated as saying that the ${\cal E}_i$ are
eigenvectors of the ``total angular momentum'' operators, with
eigenvalue zero, taking the ad-action of the ${\cal M}_j$ as a
realization of (the negative of) the usual `spin'-operators
for {\bf sl}(2,\complex).
Such a point of view ought, it seems, to generate
``nice'' expressions for the solutions; however, we have not yet seen them.
\vs
\leftline{{\bf References:}}\par
\parindent=0pt
\baselineskip=12pt
\def\hi{\hangindent=20truept}

\def\IT#1{{\it #1 \/}}
\frenchspacing
\smaller
\hi ~1.  Elie Cartan, {\it Les syst\`emes diff\'erentiels ext\'erieurs et
leurs applications g\'eom\'etriques} (Hermann, Paris, 1945).

\hi ~2. Frank Estabrook \& Hugo Wahlquist,
 J. Math. Phys. {\bf 16}, 1-7 (1975),
 {\bf 17}, 1293-7 (1976).

\hi ~3.  F. Pirani, D. Robinson and W. Shadwick, {\it Local Jet Bundle
Formulation
of B\"acklund Transformations\/}, Mathematical Physics Studies, Vol. 1 (Reidel,
Dordrecht, 1979).

\hi ~4.  Some problems are more convenient if one uses a
non-holonomic basis on the tangent space, such that
$[{\bf e}_a\,,\,{\bf e}_b] = {\bf C}^c{}_{[ab]}\,{\bf e}_c$.
Fiber coordinates are then defined by the scheme
$u_a \equiv {\bf e}_a(u),\; u_{ab} \equiv{\bf e}_b({\bf e}_a(u))$
etc., with a unique choice of order for the indices.
This behavior naturally
lifts to the corresponding total derivatives, $[D_a\,,\,D_b] =
\b C^c_{[ab]}D_c$, so that the
zero-curvature equations take the form
$D_{[a}({\bf X}_{b]}) + [{\bf X}_a,{\bf X}_b] = \b C^c{}_{[ab]}{\bf X}_c$.

\hi ~5.  I. S. Krasil'shchik \& A. M. Vinogradov,
 Acta Applic. Mathematicae {\bf 2},  79-96 (1984).

\hi ~6.  I. S. Krasil'shchik \& A.M. Vinogradov,  Acta
Applic. Mathematicae {\bf 15}, 161-209 (1989).

\hi ~7.  V.E. Zakharov and A.B. Shabat,
J.E.T.P. {\bf{34}}, 62-9 (1972),
Functional Analysis \& Applications
{\bf8}, 43-53 (1974), and {\bf13} 13-22 (1976).

\hi ~8.  A. N. Leznov, M. V. Saveliev, {\it Group-Theoretical Methods for
Integration of Nonlinear Dynamical Systems} (Birkh\"auser Verlag, Basel, 1992).

\hi ~9. J. D. Finley, III and John K. McIver,
Acta Applicandae Mathematicae {\bf32}, 197-225 (1993).

\hi 10.  C. Hoenselaers, Progress Theor. Physics \b{75}, 1014-29 (1986),
J. Phys. A \b{21}, 17-31 (1988).

\hi 11. H.-H. Chen, Phys. Rev. Lett. \b{33}, 925-8 (1974).

\hi 12. H.N. van Eck,
Proc. Kon. Ned. Akad. Wetensch, Ser. A {\bf 86}, 149-164 and
165-172  (1983).

\hi 13.  Frank B. Estabrook, in {\it Partially Integrable Evolution
Equations in Physics,} edited by R. Conte and N. Boccara (Kluwer Academic
Pubs.,
Netherlands, 1990), p. 413-434.

\hi 14. Cornelius
Hoenselaers \& Wolfgang K. Schief,
 J. Phys. {\bf A25}, 601-22 (1992).

\hi 15.  C. Hoenselaers,
Progress Theor. Physics \b{74}, 645-654 (1985).

\hi 16.  Minoru Omote, J. Math. Phys. \b{27}, 2853-60 (1986).

\hi 17.  P.H.M. Kersten, {\it Infinitesimal symmetries:  a computational
approach} (Centrum voor Wis\-kunde en Informatica, Centre for Mathematics
and Computer Science, 1009 AB Amsterdam, The Netherlands, 1987).
P.K.H. Gragert, P.H.M. Kersten, and R. Martini,
Acta Applic. Mathematicae \b{1}, 43-77 (1983).
J.H.B. Nijhof and G.H.M. Roelofs,
J. Phys. {\bf A25}, 2403-16 (1992),
 G. H. M. Roelofs and R. Martini,
J. Phys. {\bf A23}, 1877-84 (1990).

\hi 18.  C. Rogers and W. F. Shadwick, {\it B\"acklund Transformations and
Their Applications\/} (Academic Press, New York, 1982).

\hi 19.  James P. Corones and Frank J. Testa, in {\it B\"acklund
Transformations} Lecture Notes in Mathematics,
Vol. 515 (Springer-Verlag, New York, 1976)

\hi 20. Ch. Riquier, Acta Math. {\bf 25} (1902) 297-358.
E. Vessiot,
Bull. Soc. Math. France {\bf 52} (1924) 336-395;
M. Janet, {\it Le\c cons sur les syst\`emes d'\'equations aux
d\'eriv\'ees partielles}, Gauthier-Villars, Paris, 1929.

\hi 21.  M. J. Ablowitz \& H. Segur, {\it Solitons and the Inverse Scattering
Transform}, SIAM, Phil., PA, 1981, p. 171.

\hi 22.  J. D. Finley, III and John K.
McIver,
J. Math. Phys., {\bf 36}, 5707-5734 (1995).

\hi 23. R.K. Dodd and J. D. Gibbon,  Proc. R. Soc. Lond.
A. \b{359}, 411-433 (1978).

\hi 24.  W.F. Shadwick, J. Math. Phys. {\bf 19}, 2312-2317 (1978).

\hi 25.  Ulf Grenander and Gabor Szeg\"o, {\it Toeplitz Forms and Their
Applications} (University of California Press, Berkeley, Calif., 1958),
describes
earlier work in this area.
R. G. Douglas, \IT{Banach algebra techniques in the theory of
Toeplitz operators,} Conference Board of the Mathematical
Sciences, No. 15 (American Mathematical Society, Providence, RI, 1973), and
Albrecht B\"ottcher and Bernd Silbermann, \IT{Analysis of Toeplitz
Operators} (Springer-Verlag, Berlin, 1990).
are modern presentations.

\hi 26.  J. D. Finley, III, ``The Robinson-Trautman Type III Prolongation
Structure Contains K${}_{\bf 2}$,'' Commun. Math. Phys., to be published.

\hi 27.  V. G. Kac, Math. Izvestija {\bf 2}, 1271-1311 (1968),
 Bull. Amer. Math. Soc. {\bf 2}, 311-4 (1980).

\hi 28.   I. M. Gel'fand and A. A. Kirillov,
\'Etudes Sci. Publ. Math. {\bf 31}, 5-19 (1966).

\hi 29.  A. Kirillov, Contemp. Math. {\bf 145}, 1-63 (1993), p. 60.
\vfill\eject
\bye